\documentclass[dvipdfmx]{akari2017}

\usepackage{graphicx}

\shorttitle{}
\shortauthors{}

\begin{document}

\title{North Ecliptic Pole multi-wavelength survey : new optical data with Hyper Suprime-Cam and near-future prospects with eROSITA} 



\correspondingauthor{Nagisa Oi}
\email{nagisaoi@rs.tus.ac.jp}

\author{Nagisa Oi} 
\affiliation{Tokyo University of Science, 1-3 Kagurazaka, Shinjuku-ku, Tokyo 162-8601, Japan} 

\author{Hideo Matsuhara} 
\affiliation{Institute of Space and Astronautical Science, Japan Aerospace Exploration Agency, 3-1-1 Yoshinodai, Chuo-ku, Sagamihara, Kanagawa 252-5210, Japan} 

\author{Tomotsugu Goto} 
\affiliation{Institute of Astronomy, National Tsing Hua University, No. 101, Section 2, Kuang-Fu Road, Hsinchu, Taiwan} 

\author{Yousuke Utsumi} 
\affiliation{Hiroshima Astrophysical Science Center, Hiroshima University, 1-3-1 Kagamiyama, Higashi-Hiroshima, Hiroshima, 739-8526, Japan} 

\author{Rieko Momose} 
\affiliation{Institute of Astronomy, National Tsing Hua University, No. 101, Section 2, Kuang-Fu Road, Hsinchu, Taiwan} 

\author{Ting-Chi Huang} 
\affiliation{Department of Physics, National Tsing Hua University, No. 101, Section 2, Kuang-Fu Road, Hsinchu City 30013, Taiwan} 

\author{Yoshiki Toba} 
\affiliation{Academia Sinica Institute of Astronomy and Astrophysics, P.O. Box 23-141, Taipei 10617, Taiwan} 

\author{Myungshin Im} 
\affiliation{Department of Physics \& Astronomy, FPRD, Seoul National University, Shillim-Dong, Kwanak-Gu, Seoul 151-742, Korea} 

\author{Hyung Mok Lee} 
\affiliation{Department of Physics \& Astronomy, FPRD, Seoul National University, Shillim-Dong, Kwanak-Gu, Seoul 151-742, Korea} 

\author{Seong Jin Kim} 
\affiliation{Department of Physics \& Astronomy, FPRD, Seoul National University, Shillim-Dong, Kwanak-Gu, Seoul 151-742, Korea} 

\author{Takamitsu Miyaji} 
\affiliation{Instituto de Astronom\'{i}a, Universidad Nacional Aut\'{o}noma de M\'{e}xico, Ensenada, Baja California, Mexico} 

\author{Mirko Krumpe} 
\affiliation{Leibniz-Institut fur Astrophysik Potsdam, An der Sternwarte 16, 14482 Potsdam, Garmany} 

\author{Kazumi Murata} 
\affiliation{Institute of Space and Astronautical Science, Japan Aerospace Exploration Agency, 3-1-1 Yoshinodai, Chuo, Sagamihara, Kanagawa 252-5210, Japan} 

\author{Youichi Ohyama} 
\affiliation{Academia Sinica Institute of Astronomy and Astrophysics, P.O. Box 23-141, Taipei 10617, Taiwan} 

\author{Steve Serjeant} 
\affiliation{Department of Physical Sciences, The Open University, Milton Keynes, MK7 6AA, UK} 

\author{Chris Pearson} 
\affiliation{Department of Physical Sciences, The Open University, Milton Keynes, MK7 6AA, UK} 

\author{Takao Nakagawa} 
\affiliation{Institute of Space and Astronautical Science, Japan Aerospace Exploration Agency, 3-1-1 Yoshinodai, Chuo, Sagamihara, Kanagawa 252-5210, Japan} 

\author{Takehiko Wada} 
\affiliation{Institute of Space and Astronautical Science, Japan Aerospace Exploration Agency, 3-1-1 Yoshinodai, Chuo, Sagamihara, Kanagawa 252-5210, Japan} 

\author{Toshinobu Takagi} 
\affiliation{Japan Space Forum, 3-2-1, Kandasurugadai, Chiyoda-ku, Tokyo 101-0062 Japan} 

\author{Shuji Matsuura} 
\affiliation{School of Science and Technology, Kwansei Gakuin University, Sanda, Hyogo 669-1337, Japan} 

\author{Ayano Shogaki} 
\affiliation{School of Science and Technology, Kwansei Gakuin University, Sanda, Hyogo 669-1337, Japan} 

\author{NEP team} 
\affiliation{Institute of Space and Astronautical Science, JAXA, Sagamihara, Kanagawa 252-5210, Japan} 



\begin{abstract}
The AKARI North Ecliptic Pole (NEP) survey consists of two survey projects: NEP-Deep (0.5 sq.deg) and NEP-Wide (5.4 sq.deg), providing with tens of thousands of galaxies. A continuous filter coverage in the mid-infrared wavelengths (7, 9, 11, 15, 18 and 24~$\mu$m) is unique to diagnose the contributions from dusty star-formation activity and AGNs. Here we present current status focused on the newly obtained optical images and near-future prospects with a new X-ray telescope.

Hyper Suprime-Cam on Subaru telescope is a gigantic optical camera with huge Field of View (FoV).
Thanks to the wide FoV, we successfully obtained deep optical images at $g$, $r$, $i$, $z$ and $Y$-bands covering most of the NEP-Wide field.
Using the deep optical images, we identified over 5000 optical counterparts of the Mid-IR sources, presumably deeply obscured
galaxies in NEP-Wide field. 
We also investigated properties of these infrared sources with SED-fitting.

eROSITA, to be launched early 2018, is a new all-sky X-ray survey telescope, and expected to conduct ultra deep 2-10 keV imaging toward NEP. We expect unprecedentedly numerous Compton-thick AGN candidates when combined with the multi-wavelength data in NEP region.

\end{abstract}



\setcounter{page}{1}



\section{Introduction} \label{sec:intro} 
In order to understand the cosmic star-formation history, studying star-formation properties, growth of super-massive black holes (SMBHs) sitting at the central region of galaxies, their relationship and its evolution for infrared (IR) galaxies are essential because most of these activities in the early ($z\sim$ 1--2) universe are hidden by dust. 
Since it is difficult to investigate these activities in dusty galaxies with optical or shorter wavelength data due to the strong attenuation by dust, mid-infrared (MIR) and longer wavelength data are quite powerful tool. 
The AKARI North Ecliptic Pole (NEP) survey \citep{Matsuhara06} is the AKARI legacy survey spent more than 700 pointing observations with AKARI/IRC for aiming to understand the obscured universe up to $z\sim 2$.
An uniqueness of the NEP survey is that all AKARI/IRC 9 filters (reference wavelength of 2.4, 3.2, 4.1, 7.0, 9.0, 11.0, 15.0, 18.0 and 24.0~$\mu$m) were used for the whole observation area. 
Other successive IR satellites probed MIR wavelength have filter gaps, while the 9 filters of AKARI/IRC cover near-infrared (NIR) to MIR wavelength range continuously without any big filter gaps.
Thank to the uniqueness, we do not need to interpolate data to measure flux at a curtain wavelength in this wavelength range.

The AKARI NEP survey is made up of two survey projects; NEP-Deep \citep[0.5 sq.deg;][]{Wada08, Takagi12, Murata13} centered on $\alpha$ = 17$^{\rm h}$55$^{\rm m}$24$^{\rm s}$, $\delta$ = $+$66$^{\rm \circ}$37$^{\rm '}$32$^{\rm ''}$, of which limiting magnitude at 15~$\mu$m ($m_{\rm lim}^{15\mu{\rm m}}$), is 93~$\mu$Jy, and NEP-Wide \citep{Lee09, Kim12} surrounding the NEP-Deep region, which is shallower ($m_{\rm lim}^{15\mu{\rm m}}$ = 133~$\mu$Jy) but covers 10 times wider field (5.4 sq.deg). In order to take advantage of the unique filter set data to explore the hidden activities by dust, we observed multi-wavelength follow-up data from X-ray \citep{Krumpe15}, optical-NIR \citep{Oi14}, to Radio \citep{White10}. 
\citet{Matsuhara17} reported the status of the multi-wavelength observation up to 2014.
At that time our multi-band data mainly focused on the NEP-Deep field and we only have relatively shallow optical data \citep{Im10} and far-infrared (FIR) data \citep{Pearson17} for the NEP-Wide field.
About spectroscopic follow-up, we have had more than 3,000 spectroscopic data taken with MMT/Hectospec, WIYN/Hydra, Keck/DEIMOS, GTC/OSIRIS-MOS, and Subaru/FMOS.

We introduce our recent multi-wavelength data progresses covering almost entire the NEP-Wide field in $\S \ref{sec:multiFollowup}$. In $\S \ref{sec:science}$, we present preliminary results of classification of NEP-Wide IR sources into 7 types of galaxies and of evolution of their fraction through spectral energy distribution (SED) fitting. 
Then we summarize our future prospects for the NEP survey.

\section{Mutli-wavelength data for the NEP-Wide survey} \label{sec:multidata} \label{sec:multiFollowup}
\begin{figure*}
\gridline{\fig{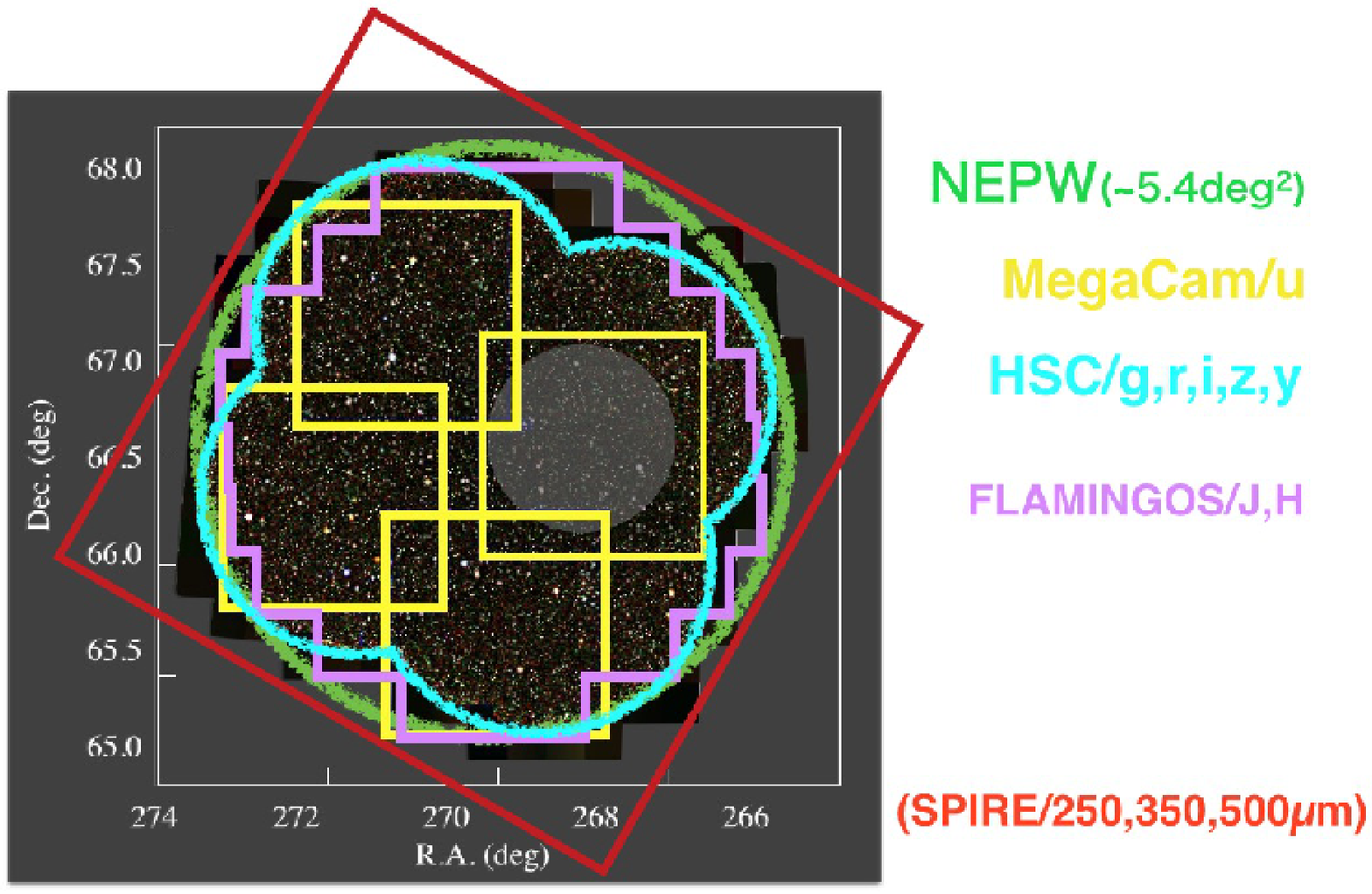}{0.4\textwidth}{}
          \fig{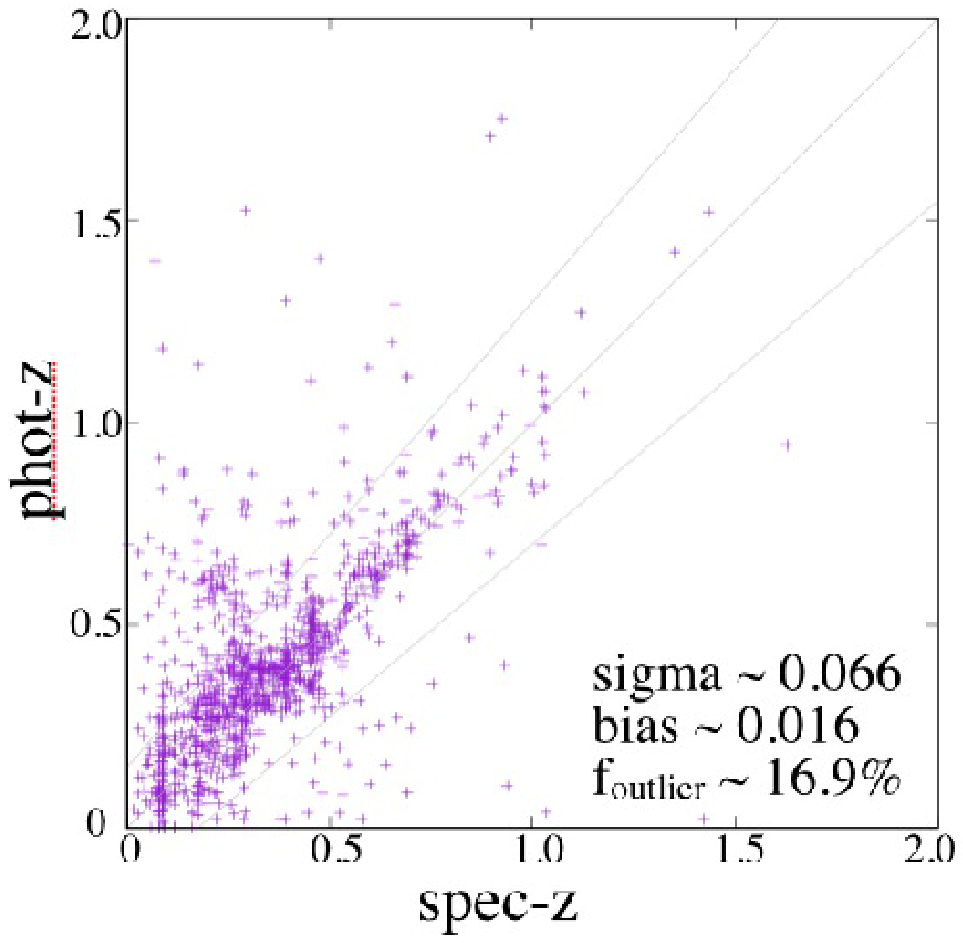}{0.25\textwidth}{}
          }
\caption{(Left): yellow squares are MegaCam $u$-band pointings (Huang et al. in prep.). HSC data ($g$, $r$, $i$, $z$, and $Y$-bands) and FLAMINGS  ($J$ and $H$-bands) observation fields are shown in cyan curve (Oi et al. in prep.) and in purple \citep{Jeon14},respectively. Red tilted square is a survey area of Herschel/SPIRE \citep[250, 350, and 500~$\mu$m;][]{Pearson17}. (Right): photometric redshift against the existing spectroscopic redshifts. The three lines represent 1:1 line and the borders of outliers, respectively.}
\label{fig:skycovzpzs}
\end{figure*}

Sky coverage of various surveys on the NEP-Wide field recently we obtained are shown in the left panel of Figure \ref{fig:skycovzpzs}.
As we can see that optical data ($u$--$Y$-bands), NIR data ($J$ and $H$) and FIR data (250, 350, and 500~$\mu$m) data covers the NEP-Wide field quite nicely.
The optical data are important to identify counterparts of NEP IR sources and calculate their photometric redshifts.
Subaru/Hyper Suprime-Cam (HSC) observation, which has a quite large FoV (1.5deg in diameter), were carried out on the NEP-Wide field spread over 6 nights on 2014 June 30, and 2015 August 7--11 (PI: T. Goto).
Thanks to the wide FoV, it can cover almost the entire NEP-Wide field with only 4 pointings. 
We reduced the data with the HSC pipeline \citep[version 4.0.1; ][]{Bosch17}, and detected more than 2million optical sources from there.
A 5 sigma limiting magnitude of each band is 27.18, 26.71, 26.10, 25.26, and 24.78 mag [AB] in $g$, $r$, $i$, $z$, and $Y$-bands, respectively.
From position matching with 3 arcsec radius search area, we found 89,178 ($\sim$89~\%) counterparts of NEP-Wide IR sources in optical data out of 100,361 sources, and the rest $\sim$10,000 IR sources are located outside of the HSC covering field.

Photometric redshifts ($z_p$) for the NEP-Wide--HSC sources were calculated using LePhare code \citep{Ilbert06} with the CFHT/MegaCam $u$-band and KPNO/FLAMINGOS $J$- and $H$-band data. 
62 galaxy models \citep{Arnouts07} and 154 stellar templates \citep{Pickles98, Bohlin95, Bixler91} are used for the fitting.
We used the existing spectroscopic redshifts for a reference of the redshift to adjust systematic offset of photometric data of each band. 
From the SED fitting, 50,210 sources are classified as galaxies and are calculated their 
The right panel of Figure \ref{fig:skycovzpzs} plots $z_p$ against the existing spectroscopic redshift ($z_s$). 
We found that $z_p$ generally agrees with $z_s$, but the dispersion is a little big.
Qualitatively, an systematic difference between $z_p$ and $z_s$ is 0.016, an accuracy of $\sigma_{\Delta z/(1+z)} \sim 0.066$ with a catastrophic error ($\sigma_{\Delta z/(1+z)}> 0.15$) rate $\eta = 16.9~\%$. 
Since the redshift range of $z_s$ covers mainly $z < 1$, the evaluation of the $z_p$ accuracy by comparing with the $z_s$ is limited to a specific range of redshift, and we need more $z_s > 1$ sample in the future to derive better quality $z_p$ and assess the accuracy.

\section{galaxy type classification and redshift evolution of fraction} \label{sec:science}
\begin{figure*}
\gridline{\fig{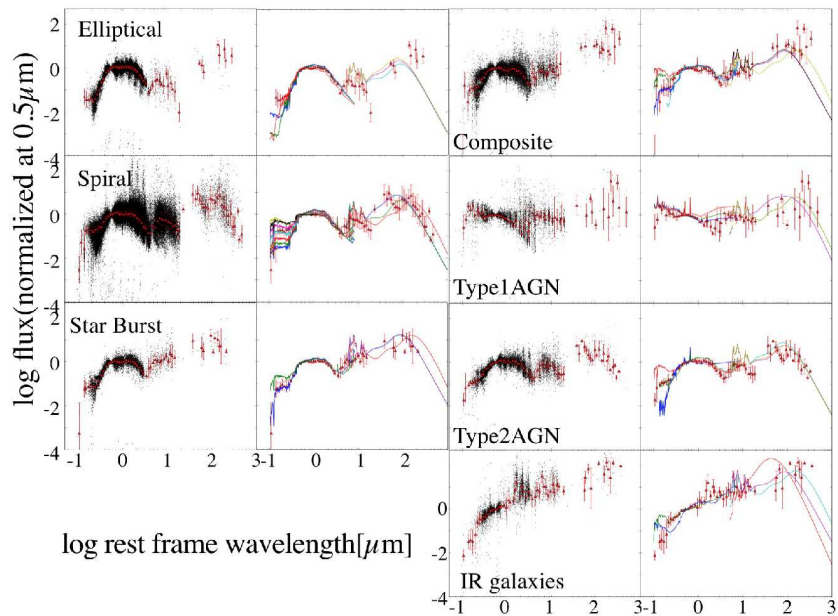}{0.643\textwidth}{}
          }
\caption{Left side panels show the rest-frame SED data (black dots) and typical SEDs (red triangles) of each galaxy type, while right ones represent the typical SEDs with galaxy modes (colored lines) used for the SED fitting.}
\label{fig:restSEDs}
\end{figure*}
There are two physical processes to radiate the MIR emission; PAH and warm dust emission from star-formation regions and emission from hot dust heated by AGNs.
Since these two emission features are totally different, our rich MIR data with wide wavelength range data enable us to distinguish them by SED fitting.
For this SED fitting, we again use LePhare code.
In this time, we adopted 24 galaxy models for shorter wavelength ($<7~\mu$m) from \citet{Polletta07} and 261 ones for longer wavelength (7 -- 1000~$\mu$m) \citep{CE01, Dale02, Lagache99}.
The first three types of models are considered as galaxies not affected by AGNs, while the rest four types of models are thought to be hosting AGNs.   
We fit these galaxy models to 24 photometric data, which are the 11 band data used for the $z_p$ calculation with AKARI/IRC eight band data (2 -- 18~$\mu$m), Herschel/PACS two bands (100 and 160~$\mu$m) and SPIRE three bands (250, 350, and 500~$\mu$m) with the fixed $z_p$. 

Rest-frame SED data points of which galaxies classified in any of the 7 types are shown in left sides of Figure \ref{fig:restSEDs}.
Median values of data with their Poisson noise per 0.05~dex normalized by the flux at 0.5~$\mu$m are over plotted.
Thank to the huge amount of galaxy sample ($\sim$50,000), typical SED shape of each galaxy type can be seen clearly.
The right side of Figure \ref{fig:restSEDs} represent the typical SEDs of each galaxy type with the models of for the classification and the top three longer wavelength models selected as the best FIR models. 
At the same time, due to the relatively small amount of FIR data, the shape of the interstellar dust thermal emission heated mainly by stars are not determined well. 
In the future, we need to have more FIR or sub-mm data to see the typical warm dust emission feature clearer.

\begin{figure*}[!t]
\gridline{\fig{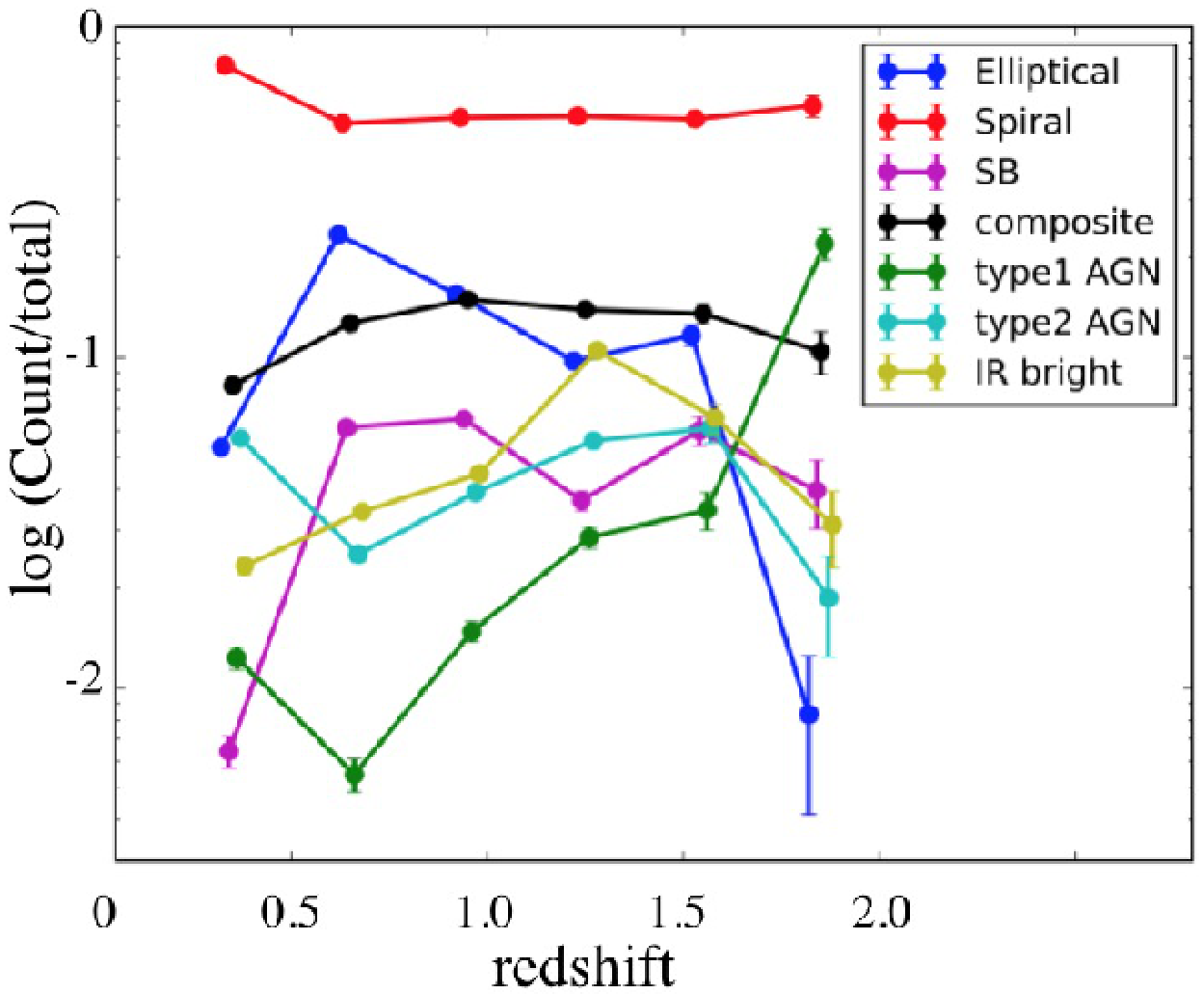}{0.35\textwidth}{}
          \fig{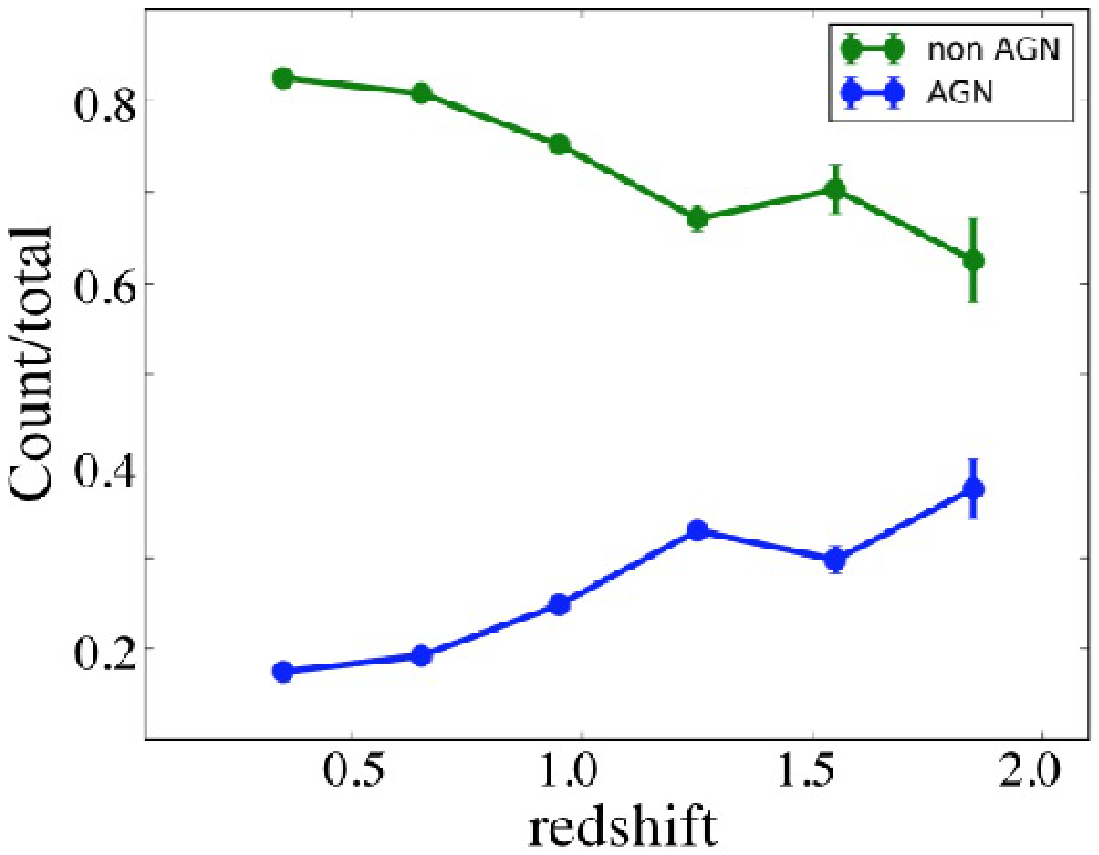}{0.35\textwidth}{}
          }
\caption{The fraction of galaxies in each type (left) and of each groups (with and without AGN) (right) as a function of redshift.}
\label{fig:fraction}
\end{figure*}

Left sides of Figure \ref{fig:fraction} show the numbers of galaxies of each type and their fractions derived from the number of each type galaxies divided by the total number of galaxies at each redshift as a function of redshift from $z =$ 0 to 2 by 0.15.
An interesting thing is that spiral galaxies account for approximately a half of our NEP-Wide IR galaxies in any redshift range.
Another interesting thing is that the fraction of type1 AGN is quite small in the local Universe, and it increases dramatically with redshift over $z > 0.5$.

We divided the 7 types of galaxies into 2 groups; hosting AGN (composite, Type1, Type2, and IR bright galaxies) and not with AGN (elliptical, spiral, and star-burst galaxies), and plotted the fractions of the galaxies in each group as a function of redshift (right side of Figure \ref{fig:fraction}).
The figure clearly shows a sense that the fraction of galaxies with AGN is gradually increasing with redshift.
However, our sample is flux-limited and some previous advocate that the fraction of AGN increases with IR-luminosity \citep[e.g., ][]{Imanishi10, Juneau13}.
Thus, the sense of AGN fraction increasing with redshift may be a bias of IR luminosity, and we have to resolve the degeneracy of redshift and IR luminosity.

\section{Near future prospects}
Up to now, we have corrected various wavelength follow-up photometric data on the NEP survey field and the existing data enable us to isolate the AGN affected galaxies from pure star-forming galaxies.
However, as we mentioned above, we need more FIR data to understand the typical SED shapes of galaxies at $z = 0-2$, and more spectroscopic data at $z>1$ to derive better photometric redshifts.
Thus, now we are planing to take new multi-wavelength survey data as below:

{\bf JCMT/SCUBA-2} -- 
The entire NEP-Wide survey field is now selected as the SCUBA2 Cosmological Legacy Survey (S2CLS) field (PI. H.-M. Lee).
This survey is expected to be held by coming March.

{\bf SRG/eROSITA} --
This is the first X-ray instrument put into the Sun-Earth L2 point onboard the Russian "Spectrum-Roentgen-Gamma" (SRG) satellite expected to launch in Spring 2018.
This will conduct 4-year all-sky survey at 0.2--12keV.
Thanks to the high visibility of NEP, we expect to find $\sim$300 Compton-Thick AGN in the NEP-Wide field.
We are discussing on possible collaboration with Russian eROSITA team.

{\bf Subaru/PFS} --
This is a new spectrograph system with a fiber positioner system to be on the Subaru telescope.
The instrument allows us to observe up to 2400 spectroscopic data simultaneously in a wide field of view of 1.25 sq.deg, covering a wavelength range of 0.38 -- 1.26~$\mu$m.
This can provide order of magnitude large amount of spectroscopic data or has a potential to derive the data for all $\sim$50,000 NEP-Wide IR sources.

\subsection*{Acknowledgments}
We are grateful to all AKARI team members for their support on this project. 
The AKARI NEP survey project activities are are partly supported by JSPS grants 16204013 and 23244040, and the Institute of Space and Astronautical Science, Japan Aerospace Exploration Agency. 
TG acknowledges the support by the Ministry of Science and Technology of Taiwan through grant 105-2112-M-007-003-MY3.



\end{document}